\documentclass{ws-ijmpd}
\usepackage[super,compress]{cite}

\begin{document}
	
\title{3D STUDY OF CENTRIFUGAL ACCELERATION IN ISOTROPIC PHOTON FIELDS
}

\author{G. G. BAKHTADZE
}

\address{School of Physics, Free University of Tbilisi\\
	Tbilisi, 0159, Georgia\\
	gbakh15@freeuni.edu.ge}

\author{V. I. BEREZHIANI}

\address{School of Physics, Free University of Tbilisi\\
	Tbilisi, 0159, Georgia
	\\
	Andronikashvili Institute of Physics (TSU), Tbilisi 0177, Georgia
	\\
	v.berezhiani@freeuni.edu.ge}

\author{Z. OSMANOV}

\address{School of Physics, Free University of Tbilisi\\
	Tbilisi, 0159, Georgia\\
	z.osmanov@freeuni.edu.ge
}

\maketitle

\begin{history}
	\received{Day Month Year}
	\revised{Day Month Year}
\end{history}

\begin{abstract}
In this paper we study relativistic dynamics of charged particles co-rotating with prescribed trajectories, having the shape of dipolar magnetic field lines. In particular, we consider the role of the drag force caused by the photon field the forming of  equilibrium positions of the charged particles. Alongside a single particle approach we also study behaviour of ensemble of particles in the context of stable positions. As we have shown, together they create surfaces where particles are at stable equilibrium positions. In this paper we examine these shapes and study parameters they depend on. It has been found that under certain conditions there are two distinct surfaces with stable equilibrium positions. 

\end{abstract}


\ccode{PACS numbers: 98.70.Sa; 97.60.Lf; 98.62.Js}

\section{Introduction} \label{sec:intro}

Recent studies of the origin of cosmic rays (Refs. \refcite{hess1,hess2}) is a long standing problem in astrophysics and therefore, understanding the origin of very high energy (VHE) particles is one of the major tasks one has to addrress. There are several proposed processes which try to explain how VHE particles have such huge energies. One of them is the so called Fermi acceleration mechanism which states that the particles are accelerated by means of collision against moving magnetic walls in interstellar medium (Refs. \refcite{fermi,bell1,bell2}), but as it has been realised, this process requires relativistic pre-acceleration of particles (Ref. \refcite{rieger}). Another approach, which we are considering in this article, is a mechanism hypothesised by Gold  for explaining high energy emission from pulsars (Refs. \refcite{gold1,gold2}). The author states that, since magnetic field in the pulsar magnetosphere is huge, charged particles follow the rapidly co-rotating field lines and as a result they experience extremely strong centrifugal force, leading to generation of VHE particles.

If one takes relativistic effects into account, some interesting phenomena can happen. In particular, in (Ref. \refcite{mr}) authors have studied dynamics of particles moving along co-rotating straight channels. As it has been shown, particles initially radially accelerate, but since azimuthal velocity increases by reaching the light cylinder (LC) surface (a hypothetical area, where the linear velocity of rotation exactly equals the speed of light), in due course of time, the radial component starts decreasing. It is worth noting that VHE particles observed nearby the Solar system are good indicators of force-free dynamics, which in turn, means that a good approximation to describe such a system is to assume that particles follow co-rotating field lines having the shape of Archimede's spiral (Ref. \refcite{rdo}). Mechanism of magnetocentrifugal acceleration has been successfully applied to a variety of astrophysical objects such as black holes (Refs. \refcite{neutr,pevatron,newmech,applic5}) and pulsars (Refs. \refcite{or17,group,applic2,or09}). 

Generally speaking, it is worth noting that magnetospheres of astrophysical objects are imbedded in photon fields, the origin of which might be either thermal or nonthermal radiation. Therefore, particle dynamics might be strongly influenced by existence of the mentioned fields. In particular, the photon field will inevitably create a drag force, which will alter the overall dynamics of particles. In (Ref. \refcite{me}) we have considered this problem for the co-rotating channels located in the equatorial plane. It has been found that by means of the photon drag force the particle acceleration is slightly reduced. One of the interesting features, characterising such systems, is that under certain conditions particles gather at stable equilibrium positions. These areas, in principle, might be detectable and therefore, it is even necessary to generalise the previous work and study the same problem for three-dimensional configurations of magnetic field lines and see the 3D map of such equilibrium zones.

The paper is organised in the following way. In section 2 we present a theoretical model of centrifugal acceleration in the background of isotropic photon field, in section 3 we consider our results and in section 4 we summarise them.

\section{Theoretical model} \label{sec:main}

In this section we examine the motion of particles along channels rotating with constant angular velocity, $\omega$, around $z$ axis. The shape of the channel-wire is given parametrically (with the parameter $p$) in Cartesian coordinates:
\begin{equation}\label{shape1}
  \mathbf{r} = x(p) \mathbf{\hat{i}} + y(p) \mathbf{\hat{j}} + z(p) \mathbf{\hat{k}}
\end{equation}
As in the previous paper (henceforth paper-I) we study dynamics of centrifugally accelerated particles influenced by the relatively dense photon field, which are normally present in spinning magnetospheres of some astrophysical objects. In the framework of the paper we assume that the "photon sea" is isotropic in the laboratory frame (LF) of reference. By means of the photons, the particles apart from the centrifugal effects experience the so-called photon drag force 
\begin{equation}\label{drag}
  \mathbf{F_f} = - \beta \gamma^2 \mathbf{V},
\end{equation}
where by $\gamma$ we denote the Lorenz factor of the particle, $\mathbf{V}$ denotes its velocity in the laboratory reference frame (LF), $\beta$ is presented by 
\begin{equation}\label{beta}
 \beta = \frac{4}{3}\sigma_{_T}U
\end{equation}
and $U$ is the energy density of the photon field and for the speed of light we use $c=1$.

For the rotating system the radius vector of the particle in the LF at any moment of time can be given by
\begin{equation}\label{shape}
  \mathbf{R} = X \mathbf{\hat{i}} + Y \mathbf{\hat{j}} + Z \mathbf{\hat{k}},
\end{equation}
where
\begin{eqnarray}
  X &=& x(p(t)) \cos{w t} - y(p(t)) \sin{w t},\\
  Y &=& x(p(t)) \sin{w t} + y(p(t)) \cos{w t},\\
  Z &=& z(p(t)).
\end{eqnarray}
Consequently, the total velocity of particle in the LF writes as 
\begin{equation}
  \mathbf{V} = \frac{d \mathbf{R}}{dt}.
\end{equation}
Likewise the approach developed in paper-I, effectively the particles sliding along prescribed corotating trajectories with constant shape in the corotating frame (CF) of reference, except $\mathbf{F_f}$ also experience the reaction force, $\mathbf{N}$. Therefore, the equation of motion is given by
\begin{equation}
\label{eqmo}
  \frac{d \mathbf{P}}{dt} = \mathbf{F_f} + \mathbf{N}
\end{equation}
where $\mathbf{P}$ denotes Particle's relativistic momentum
\begin{equation}
  \mathbf{P} = \gamma m \mathbf{V},
\end{equation}
and $m$ is the mass of the particle.

By introducing the tangent vector to the path of the particle
\begin{equation}\label{a}
  \mathbf{T} = \frac{\partial \mathbf{R}}{\partial p},
\end{equation}
from Eq. (\ref{eqmo}) one can derive following identity
\begin{equation}\label{b}
  \left(\frac{d \mathbf{P}}{dt} - \mathbf{F_f}\right) \cdot \frac{\partial \mathbf{R}}{\partial p} = 0,
\end{equation}
which after straightforward simplifications lead to the following second order ordinary differential equation for $p(t)$

\begin{figure}
	\centering
	\includegraphics[width=\textwidth]{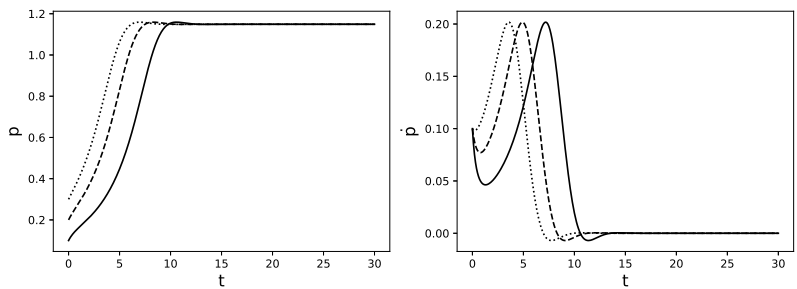}
	\caption{Here we plot graphs of $p$ and $\dot{p}$ versus time for different initial positions. The set of parameters is $m=1$, $\omega = 1 $, $\beta = 1.0$, $L=1$, $\alpha=\pi/3$ and $\phi=\pi/4$, $\dot{p}(0)=0.1$, $p(0)=0.3$ (dotted line), $p(0)=0.2$ (dashed line) and $p(0)=0.1$ (solid line). }\label{difps}
\end{figure}

\begin{equation}
\label{d2p}
\frac{d^2 p}{dt^2}=\frac{\beta \left(\dot{p} S+ \omega d \right) \sqrt{\Delta}}{m
	\Gamma}+\frac{P_3 \dot{p}^3 + P_2 \dot{p}^2 + P_1 \dot{p} +P_0}{\Gamma}
\end{equation}

where all supplementary functions, $d$, $r$, $S$, $S_p$, $r_p$, $\Gamma$, $\Delta$, $P_0$, $P_1$, $P_2$, $P_3$, $P_x$ and $P_y$ are explicitly given in Appendix A (see Eqs. {\ref{apend}}).
It is obvious that a solution for the parameter $p(t)$ is enough to describe dynamics of the particle because it moves alongside a prescribed channel.

\section{Discussion} \label{sec:res}
%
%
%
%

In this section we study stable equilibrium states which are caused by photon field drag force. For this purpose, as a natural example, we examine dipolar configuration of magnetic field lines. We assume that the magnetic dipolar momentum is inclined by the angle $\phi$ with respect to the axis of rotation. The corresponding shape of the field lines in the XOY plane is given parametrically as

\begin{eqnarray}
\label{dipfield}
x_0 &=& L \sin^2{p} \cos{p},\\
y_0 &=& L \sin^3{p},
\end{eqnarray}
where $L$ defines the length-scale of the dipolar field line. Any other field line might be given by the rotation of the curve with the angle $\alpha$ around the $x$ axis and with $\pi/2-\phi$, around the $y$ axis respectively, resulting in the following parametric form of the curve 
\begin{eqnarray}
\label{3dipfield}
x &=& x_0 \sin{\phi} + y_0 \sin{\alpha} \cos{\phi},\\
y &=& y_0 \cos{\alpha},\\
z &=& y_0 \sin{\phi} \sin{\alpha} - x_0 \cos{\phi},
\end{eqnarray}
which should complement Eq.~(\ref{d2p}).

\begin{figure}
	\centering
	\includegraphics[width=\textwidth]{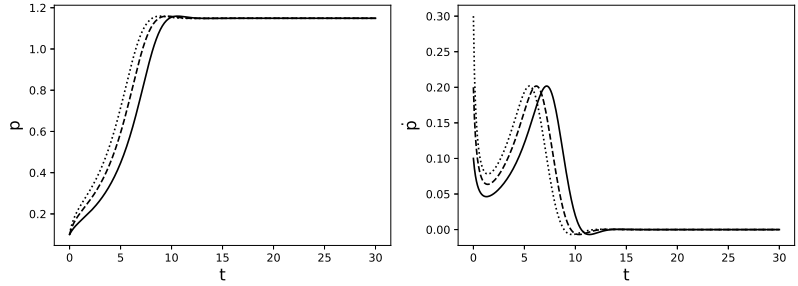}
	\caption{Plots of $p(t)$ and $\dot{p}(t)$ with different values of $\dot{p}(0)$. The set of parameters is $m=1$, $\omega = 1 $, $\beta = 1.0$, $L=1$, $\alpha=\pi/3$ and $\phi=\pi/4$, $p(0)=0.1$, $\dot{p}(0)=0.3$ (dotted line), 
$\dot{p}(0)=0.2$ (dashed line) and $\dot{p}(0)=0.1$ (solid line). }\label{difDps}
\end{figure}
\begin{figure}
	\centering
	\includegraphics[width=\textwidth]{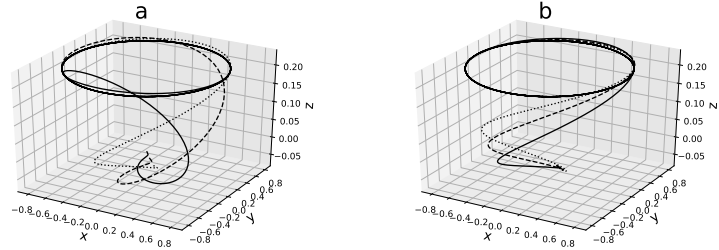}
	\caption{(a) Trajectories in the LF for different initial positions. The set of parameters are $m=1$, $\omega = 1 $, $\beta = 1.0$, $L=1$, $\alpha=\pi/3$ and $\phi=\pi/4$, $\dot{p}(0)=0.1$, $p(0)=0.3$ (dotted line), $p(0)=0.2$ (dashed line) and $p(0)=0.1$ (solid line). (b) Trajectories in the LF for different values of $\dot{p}(0)$. The set of parameters are $m=1$, $\omega = 1 $, $\beta = 1.0$, $L=1$, $\alpha=\pi/3$ and $\phi=\pi/4$, $p(0)=0.1$, $\dot{p}(0)=0.3$ (dotted line), $\dot{p}(0)=0.2$ (dashed line) and $\dot{p}(0)=0.1$ (solid line).  }\label{3dtr}
\end{figure}

\subsection{Motion of one particle}

In paper-I we have shown that the particles following the field lines either will gather somewhere inside 
the LC because of the photon drag force, or reach the area of the LC. In the latter case the effective reaction force
becomes extremely large and the field lines cannot to hold the particle anymore.

We consider several interesting cases with different parameters. In Fig.~\ref{difps} we show the behaviour of
$p(t)$ and $\dot{p}(t)$ for different values of ${p}(0)$. The set of parameters is $m=1$, $\omega = 1 $, $\beta = 1.0$, $L=1$, $\alpha=\pi/3$ and $\phi=\pi/4$, $\dot{p}(0)=0.1$, $p(0)=0.3$ (dotted line), $p(0)=0.2$ (dashed line) and $p(0)=0.1$ (solid line).  As it is clear from the plots, as $p(t)$ as $\dot{p}(t)$ asymptotically tend to equilibrium values. On the other hand, these values explicitly define location and velocity of a particle, therefore  one can conclude that a stable equilibrium position is independent of initial conditions. This is a general feature and can be straightforwardly checked by direct numerical calculations. In particular, in Fig.~\ref{difDps} we plot the similar graphs, but for different values of $\dot{p}(0)$. The set of parameters is $m=1$, $\omega = 1 $, $\beta = 1.0$, $L=1$, $\alpha=\pi/3$ and $\phi=\pi/4$, $p(0)=0.1$, $\dot{p}(0)=0.3$ (dotted line), $\dot{p}(0)=0.2$ (dashed line) and $\dot{p}(0)=0.1$ (solid line). Likewise the results shown in the previous figure, it is clear that in due course of time the values of $p(t)$ and $\dot{p}(t)$ both tend to asymptotic values, which means that the particles reach an equilibrium position.
\begin{figure}
	\centering
	\includegraphics[width=\textwidth]{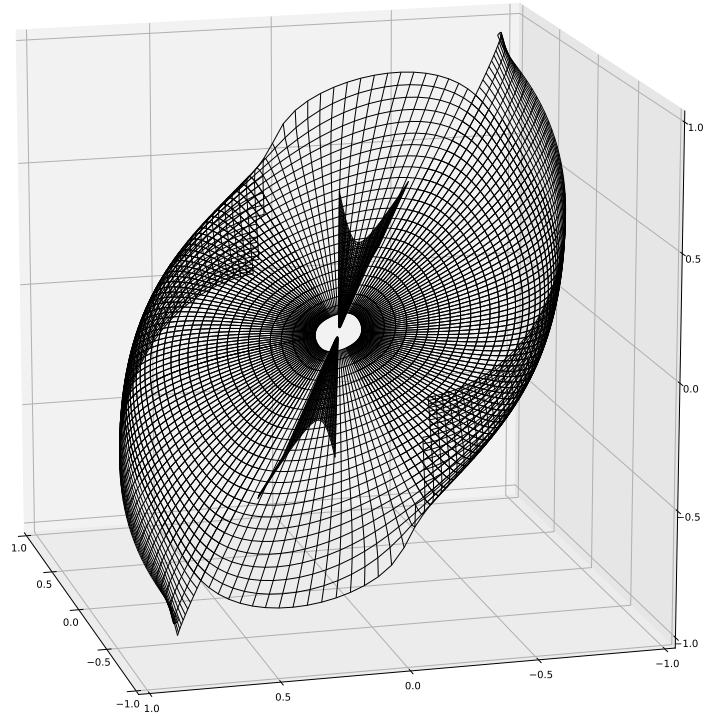}
	\caption{Plot for the stable equilibrium plane. The set of parameters are $m=1$, $\omega = 1 $, $\beta = 1.0$ and $\phi=\pi/4$.}\label{3dprev}
\end{figure}

The asymptotic value of $p$ means that the particles's position on the field line is constant. On the other hand, the field lines are co-rotating, which means that asymptotically the trajectory of particles in the LF will tend to a circular shape. 

\begin{figure}
	\centering
	\includegraphics[width=\textwidth]{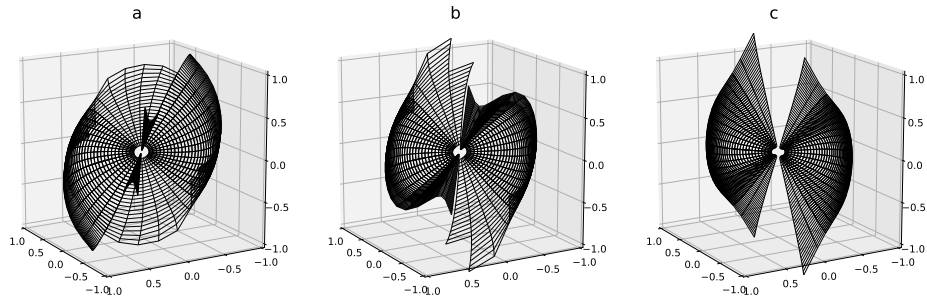}
	\caption{Plot for the stable equilibrium surfaces for different values of $\phi$. The set of parameters is $m=1$, $\omega = 1 $, $\beta = 1.0$, (a) $\phi=\frac{\pi}{4}$, (b) $\phi=\frac{\pi}{3}$, (c) $\phi=\frac{\pi}{2}$.}\label{3dty}
\end{figure}

In Fig.~(\ref{3dtr}) we show the trajectories of particles in the LF of reference. In particular, on (a) we plot the trajectories of particles corresponding to the set of parameters of Fig.~\ref{difps} and on (b) we demonstrate the results corresponding to Fig.~\ref{difDps}. As it is clear from both figures, regardless of the initial conditions the particles's trajectories, initially having a shape of spirals, asymptotically tend to circular motion. 

\subsection{Stable equilibrium positions}

As we have already seen, a single field line collects the particles in certain positions. Other field lines might have equilibrium positions in different points (Ref. \refcite{me}). In this context it is worth noting that each field line  is characterised by two parameters $L$ and $\alpha$, which in turn means that stable equilibrium points of all field lines will create a surface. This means that from the co-rotating frame (CF) of reference the charged particles will gather on a certain two dimensional surface as shown in Fig.~\ref{3dprev}. The set of parameters is $m=1$, $\omega = 1 $, $\beta = 1.0$ and $\phi=\pi/4$. As it is evident, for the mentioned parameters, independently on the initial conditions the particles sliding along different magnetic field lines finally will be gathered on a surface with a quite non trivial shape. On the other hand, the shape itself should depend on central object's angular velocity, $\omega$, angle between the angular velocity vector and the dipole momentum, $\phi$, and the parameter $\beta/m$. 

As a first example we examine dependence on $\phi$. In Fig.~\ref{3dty} we show equilibrium surfaces for different \begin{figure}
	\centering
	\includegraphics[width=\textwidth]{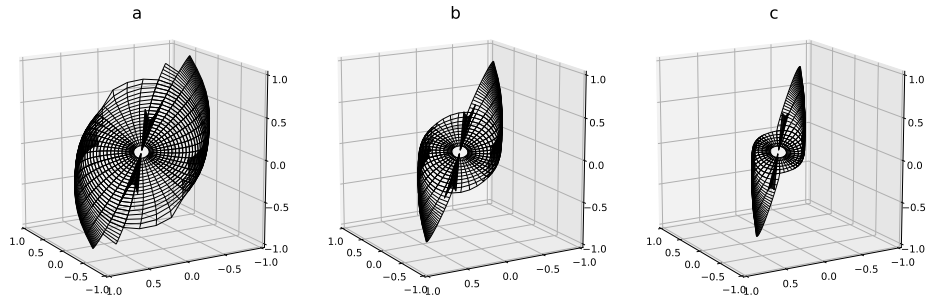}
	\caption{The stable equilibrium surfaces for different values of angular velocities. The set of parameters is $m=1$, $\phi = 1 $, $\beta = 1.0$, (a) $\omega=1.2$, (b) $\omega=2.0$, (c) $\omega=3.2$.}\label{3dw}
\end{figure}
angles between the angular velocity and the magnetic dipole momentum. The set of parameters is $m=1$, $\omega = 1 $, $\beta = 1.0$, (a) $\phi=\frac{\pi}{4}$, (b) $\phi=\frac{\pi}{3}$, (c) $\phi=\frac{\pi}{2}$. As it is clear from the plots, higher values of $\phi$ lead to the higher deviation of the surfaces from the equatorial plane. Another interesting feature is splitting of the surfaces which becomes more distinct also for higher values of the inclination angle. In particular, for $\phi = \pi/2$ (see (c) in Fig.~\ref{3dty}) one can clearly distinguish two different areas of the equilibrium surface.

In Fig.~\ref{3dw} we demonstrate the dependence of shapes of equilibrium surfaces on the angular velocity of rotation. The set of parameters is $m=1$, $\phi = 1 $, $\beta = 1.0$, (a) $\omega=1.2$, (b) $\omega=2.0$, (c) $\omega=3.2$. The plots make it evident that for higher values of $\omega$ the corresponding structure is relatively squeezed, becoming close to the axis of rotation. The reason of such a behaviour is that, on the one hand, stable equilibrium position can't exist outside of the LC area because the velocity of a particle in the LF cannot exceed the speed of light. On the other hand, the LC radius behaves as $1/\omega$, therefore, the corresponding surface will be more squeezed for higher values of angular velocities.

\begin{figure}
	\centering
	\includegraphics[width=\textwidth]{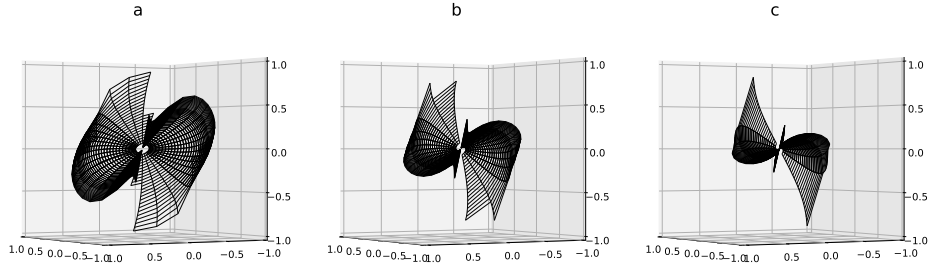}
	\caption{The stable equilibrium surfaces for different  values of $\beta/m$. The set of parameters is $m=1$, $\phi = 1 $, $\omega = 1.0$, (a) $\beta/m=1.6$, (b) $\beta/m=3.0$, (c) $\beta/m=6.0$.}\label{3dmb}
\end{figure}

The behaviour of a shape of the surface versus $\beta/m$ is shown in Fig.~\ref{3dmb}. The set of parameters is $m=1$, $\phi = 1 $, $\omega = 1.0$, (a) $\beta/m=1.6$, (b) $\beta/m=3.0$, (c) $\beta/m=6.0$. As one can see from the plots, by increasing the effects of the drag force (increasing value of $\beta/m$) more particles are gathered close to the axis of rotation. This is a natural results, because as it is clear from Eq. (\ref{drag}), the drag force is proportional to $\gamma^2 V$. On the other hand for higher values of $\beta/m$ the Lorentz factor will be smaller and consequently particles mostly will gather in a relative vicinity of the rotation axis and as a result the structure will be squeezed.

\section{Summary} \label{sec:summary}
%
%
%
%

\begin{enumerate}
     \item We considered three dimensional dynamics of particle motion in the co-rotating magnetospheres with the presence of the photon "sea" to understand the effect of the photon field by means of the radiation drag force on motion of particles. It has been shown that each of the field lines might have its own stable equilibrium position which depends on several parameters.
     \item By taking the dipolar magnetic field configuration into account we have shown that considering all field lines equilibrium points create a surface. Obviously this surface is located inside the LC, which in turn, means that the bigger the angular velocity the smaller the surface length-scales. Shape of this surface depends on different parameters such as its angular velocity of rotation, dipole momentum inclination and the ratio of drag force coefficient and the particle mass.
     \item We have found that as $\phi$ increases a stable equilibrium surface changes its whole configuration: for relatively small angles there is only one surface, which for higher values of $\phi$ splits into two different surfaces. Similar analysis with considering different values of angular velocity has revealed that the corresponding structure becomes more squeezed for higher values of $\omega$. By studying the shape of the surface versus $\beta/m$, somewhat similar results has been obtained: for larger values of the photon drag force the surface of the equilibrium state becomes closer to the rotation axis but it is still spanned to the LC area.
\end{enumerate}

In this paper we have studied the role of photon drag force theoretically, without applying the mathematical model to particular astrophysical scenarios. On the other hand, it is obvious that magnetospheres of pulsars and active galactic nuclei might provide all necessary conditions for the frozen-in condition, which is a key point for the current approach. As a next step it could be interesting to apply the developed model to the mentioned class of objects and see how the centrifugal mechanism of acceleration is reduced by means of the photon sea.
Generally speaking, the photon sea can be provided by the synchrotron emission of the same particles involved in the process of acceleration. Therefore, the tools developed in this paper might have very interesting consequences in different astrophysical scenarios.

\section*{Acknowledgments}

The research was supported by the Shota Rustaveli National Science Foundation grant (DI-2016-14) and partially by the grant (FR17-587). The research of G.B. was supported by the Knowledge Foundation at the Free University of Tbilisi. Z.O. acknowledges the hospitality of the Max-Plank Institute for Nuclear Physics (Heidelberg, Germany) during his visit in 2018 in the framework of DAAD scholarship for Research Stays for University Academics and Scientists. Z.O. is especially grateful to prof. F. Aharonian for his interesting comments.

\appendix

\section{Appendices}
Below we give explicit expressions of all supplementary functions used in Eq. (\ref{d2p})

$$ d = x y' - x' y, \;\; r^2 = x^2+y^2, \;\; S = x'^2+y'^2+z'^2, \;\;S_p = \frac{1}{2} \frac{\partial S}{\partial p} = x' x''+y' y''+z' z'',$$

$$r_p = \frac{1}{2} \frac{\partial r^2}{\partial p} = x x'+y y', \;\;\Gamma = x'^2 \left(x^2 \omega ^2-1\right)+z'^2 \left(\omega ^2 r^2-1\right)+2 x x' y \omega ^2 y'+y'^2 \left(y^2 \omega ^2-1\right),$$

$$\Delta = -\dot{p}^2 S -2 \dot{p} \omega d  - \omega ^2 r^2+1, \;\; P_0 = \omega ^2 \left(\omega ^2 r^2-1\right) r_p, \;\;P_1 = 3 \omega ^3 d r_p, $$

$$P_2 = -\omega ^2 \left(z' z'' r^2+r_p \left[x x''+y y'' -2 S\right]\right)+S_p, \;\;\;\;\;\;\;\; P_3 = \omega  (P_x+P_y), $$

\begin{equation}
\label{apend}
P_x = x \left(y'' \left[x'^2+z'^2\right] - y' \left[x' x'' + z' z''\right]\right), \;\;P_y = y \left(x' \left[y' y'' + z' z''\right] - x'' \left[y'^2 + z'^2\right]\right),
\end{equation}

where $\psi'\equiv d\psi/dp$, $\psi''\equiv d^2\psi/dp^2$, $\dot{\psi}\equiv d\psi/dt$ and by $\psi$ we denote any of the physical quantities.

\end{document}